# An Enhanced Gradient Based Optimized Controller for Load Frequency Control of a Two Area Automatic Generation Control System

Nabil Anan Orka[1][0000-0001-5251-2137], Sheikh Samit Muhaimin[1][0000-0002-7055-9024], Md. Nazmush Shakib Shahi[1][0000-0002-7209-163X] and Ashik Ahmed[1][0000-0002-8592-7563]

[1] Department of Electrical and Electronic Engineering, Islamic University of Technology, Board Bazar, Gazipur-1704, Bangladesh
nabilanan,samitmuhaimin,nazmushshakib,ashik123@iut-dhaka.edu

**Abstract.** This work proposes the adoption of Enhanced Gradient-Based Optimizer (EGBO) as a new approach to the Load Frequency Control (LFC) problem in a two-area interconnected power system. The importance of determining the optimal parameters for the controllers for the LFC problem cannot be overstated, and the fact that estimating these parameters require complex and nonlinear computations makes the optimization procedure even more unique and challenging. Consequently, application of an efficient optimization algorithm to successfully attain optimal controller parameters is critical. To accomplish this task, the proposed EGBO algorithm is compared to the fundamental Gradient-Based Optimizer (GBO), Chimp Optimization Algorithm (ChOA), Sine Cosine Algorithm (SCA), Grey Wolf Optimization (GWO), and Particle Swarm Optimization (PSO) for optimizing an Integral-Time-multiplied-Absolute-Error (ITAE) based objective function. The relevant findings show that the EGBO algorithm is competitively superior in terms of resilience, precision, and latency when compared to other optimization methods. Lastly, the statistical comparison further strengthens the outcome of the study.

## 1 Introduction

For a power generation system, the power demand is always changing, necessitating a change in the level of generated power to establish a balance between supply and demand. Due to the irregularity of the load demand, the speed of the generator is altering consequently, which alters the generated power frequency as it is directly proportional to the speed of the generator [1]. To keep the frequency variation within acceptable limits, Automatic Generation Control (AGC) is usually incorporated into the system to adjust the generation of the power system following any changes in the load. In an interconnected system, the role of AGC is to maintain reasonable frequency while dividing the load among available generating units [2].

Load Frequency Control (LFC) is the frequency control mechanism of an AGC system, integrating feedback loops of several systems connected through a tie-line. For a stable system response, the incorporation of an efficient controller is vital considering the sensitivity and flexibility of the systems. To date, Proportional-Integral-Derivative (PID) controllers are one of the vastly adopted techniques due to their simplicity of operation and cost-effectiveness. However, tuning a PID controller manually or in a trial & error method is a difficult and meticulous task. Optimization algorithms are ideal for such cases as they search for optimal parametric values for a certain objective function within specified system constraints.

In [3-12] Integral-Time-multiplied-Absolute-Error (ITAE) values are not explicitly presented or other performance metrics are used. In [4, 13-15], PID gains of only one area are optimized by the proposed controllers. In [4, 13-16], Step Load Perturbation (SLP) of only one area at a time are studied, ranging from 1 to 10 percent. Small execution times are critical in the LFC domain because loads are dynamic and optimized controllers must be able to cope up with the change. However, no surveyed literature contains the execution times of the optimization algorithms.

In this work, an ITAE based objective function is adopted to minimize the Area Control Error (ACE) of a two-area all thermal system. The overall system model is developed using the state–space approach. Later, generic PID controller blocks are incorporated to enhance the system time-domain characteristics, such as steady-state error, settling time and overshoot. Different SLPs ranging from 10% to 25% are applied to both areas as well as separate individual areas to study the robustness and flexibility of the controller. The resulting frequency deviation (FD) and tie-line power deviation (TPD) are tracked to formulate the ITAE based objective function.

For the optimization of the generic PID controller, an Enhanced Gradient-Based Optimizer (EGBO) is proposed in this work, which utilizes Newton's method to traverse the solution space [29]. It is an update of the Gradient-Based Optimizer (GBO), incorporating new operators such as a crossover operator, and a modified local escaping operator [30]. To enhance precision and convergence speed, EGBO utilizes a new mechanism to update the control parameters [29]. To the best of the authors' knowledge, the usage of EGBO in the LFC problem of multi-area systems is still unexplored. For a comprehensive analysis, widely applied nature-inspired algorithms such as PSO and GWO tested on [3, 11, 16, 22], a population-based SCA tested on [13] are used, along with recently developed meta-heuristic Chimp Optimization Algorithm (ChOA) [31] and GBO. Execution timings of the investigated algorithms are presented along with ITAE values, to demonstrate how they can function in a dynamic context of load perturbations.

The key contributions in this research can be summarized as follows:

**Table 1.** A Brief Literature Survey of Load Frequency Control

| Year | Ref. | System type | No. of areas | Controller type | Optimizer type |
|------|------|-------------|--------------|-----------------|----------------|
| 2016 | [16] | Thermal, hydro-thermal, and multi-sources | 2,3 | Proportional-Integral (PI), and PID | Grey Wolf Optimization (GWO), Comprehensive Learning, Particle Swarm Optimization (CLPSO), and Ensemble of Mutation and Crossover Strategies and Parameters in Differential Evolution (EPSDE) |
| 2016 | [4] | Thermal | 2 | PID | Seeker Optimization Algorithm (SOA) |
| 2016 | [17] | Thermal | 1 | Interval Fractional Order PID (INFOPID) | Kharitonov's theorem |
| 2017 | [18] | Islanded microgrid | - | PI, and Fuzzy PI | Modified Harmony Search Algorithm (MHSA) |
| 2018 | [13] | Thermal | 2 | PID with a derivative filter (PIDN) | Sine-Cosine Algorithm (SCA) |
| 2019 | [5] | Multi sources | 2 | PID | JAYA Algorithm (JA) |
| 2019 | [19] | Thermal, and hydro-thermal | 1,2 | Fractional Order PID (FOPID) | Moth Flame Optimization (MFO) |
| 2020 | [6] | Thermal | 1,2 | FOPID | Moth Flame Optimization (MFO) |
| 2020 | [7] | Thermal | 3 | PI-DF | Coyote Optimization Algorithm (COA) |
| 2020 | [20] | Thermal | 3 | PID | Class Topper Optimization (CTO) |
| 2020 | [21] | Hybrid | 2 | FOPID | Atom Search Optimization (ASO) |
| 2020 | [8] | Hydro-thermal | 2 | Cascaded PI-PD | Adaptive Jaya Optimization Algorithm (AJOA) |
| 2020 | [9] | Thermal | 2 | Proportional-Derivative with derivative filter cascaded with 1-PI (PDF+1PI) | MFO |
| 2020 | [22] | Thermal | 2 | High Order Differential Feedback Controller (HODFC), and Fractional HODFC (FHODFC) | Particle Swarm Optimization (PSO) |
| 2020 | [23] | Thermal | 2 | Fuzzy PID (FPID) | Improved Ant Colony Optimization (IACO) |
| 2020 | [10] | Multi sources | 3 | Cascaded Fuzzy PD-PI (FPD-PI) | Grasshopper Optimization Algorithm (GOA) |
| 2020 | [24] | Multi sources | 3 | FPID | Spotted Hyena Optimization Technique (SHOT) |
| 2021 | [11] | Thermal | 2 | FOPID | PSO |
| 2021 | [25] | Thermal | 2 | Two Degrees Of Freedom-PID (2DOF-PID) | Seagull Optimization Algorithm (SOA) |
| 2021 | [26] | Multi sources | 2 | Parallel combination of Hybrid Fractional Order and Tilt-Integral-Derivative with Filter (HybFO-TIDF) | Marine Predator Algorithm (MPA) |
| 2021 | [27] | Hybrid | 3 | PI incorporated Reinforced Learning Neural Network (PI-RLNN) | Student Psychology Optimization Algorithm (SPOA) |
| 2021 | [3] | Islanded microgrid | - | PID | PSO-based Artificial Neural Network (ANN) |
| 2021 | [14] | Thermal | 2 | PIDN | Elephant Herding Optimization (EHO) |
| 2021 | [15] | Thermal, and multi sources | 2,3 | PID | Quasi-Oppositional Dragonfly Algorithm (QODA) |
| 2022 | [12] | Thermal | 4 | Active Disturbance Rejection Controller (ADRC) | Chaotic Fractional-Order Beetle Swarm Optimization (CFBSO) |
| 2022 | [28] | Thermal | 2 | FOPID | BAT algorithm |

- To investigate the LFC, a two-area non-reheat thermal-thermal system is modelled by the state–space approach.
- To optimize PID controller gain values, EGBO is deployed, along with GBO, GWO, PSO, SCA, and ChOA, for a comprehensive comparative analysis.
- To examine the PID controller's robustness and adaptability, multiple SLP ranging from 10% to 25% are applied to both regions as well as isolated individual areas.
- To further highlight the results, a rigorous statistical study is conducted, consisting of Friedman test and Wilcoxon Signed-Rank test.

The subsequent sections of the papers contain a brief discussion and modeling of the system, overview of different optimization algorithms, implementation of EGBO in the LFC problem, discussion of the simulated results of the proposed five cases with a detailed statistical analysis, and finally, the conclusion of the study.

## 2    Mathematical Model of the System

LFC problem persists in single area systems and multi-area systems, as shown in [3–28], incorporating reheated thermal systems, non-reheated thermal systems, hydro-thermal power systems, hybrid and multiple sources power systems, microgrids, etc. In this work, a two-area AGC system is studied, which is adopted from [32].

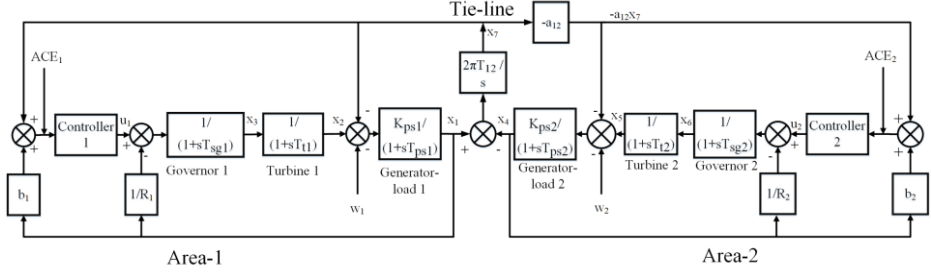

**Fig. 1.** State-space model of the system under study

Both areas in the AGC system are equipped with identical non-reheat thermal power stations. The mathematical models for different components of the adopted thermal power station are discussed in the following subsections.

### 2.1 Governor Model

Governor is responsible for operating valves to control steam flow into the turbine according to the feedback and controller output of the system. The transfer function of the governor is as follows [32]:

$$G_{sg}(s) = \frac{1}{1+sT_{sg}} \tag{1}$$

### 2.2 Turbine Model

The turbine is driven by highly pressurized steam which is responsible for driving the generator shaft. The first-order transfer function of steam turbine is as follows [32]:

$$G_t(s) = \frac{1}{1+sT_t} \tag{2}$$

### 2.3 Generator-Load Model

When the steam turbine rotates, the shaft of the generator also rotates, resulting in the generator output. The simplified transfer function of the generator-load model is as follows [32]:

$$G_{ps}(s) = \frac{K_{ps}}{1+sT_{ps}} \tag{3}$$

## 2.4　Tie-Line

Tie-lines are the transmission links that connect one area to its neighboring regions. As the name implies, LFC manages power flow between different interconnected areas while maintaining a constant frequency. To achieve 0% steady-state error in tie-line power flow, an integral controller is used in this work [32].

Detail of the nomenclatures used in the system under study can be found in appendix A. The nominal parametric values of the system, which are selected for simulations, are provided in appendix B.

The state-space model of the PID incorporated two-area thermal power system is represented as:

$$\dot{x}_1 = \frac{K_{ps1}}{T_{ps1}}(x_2 - x_7 - w_1) - \frac{1}{T_{ps1}}x_1 \tag{4}$$

$$\dot{x}_2 = \frac{1}{T_{t1}}(x_3 - x_2) \tag{5}$$

$$\dot{x}_3 = \frac{1}{T_{sg1}}(u_1 - \frac{x_1}{R_1} - x_3) \tag{6}$$

$$\dot{x}_4 = \frac{K_{ps2}}{T_{ps2}}(x_5 - a_{12}x_7 - w_2) - \frac{1}{T_{ps2}}x_4 \tag{7}$$

$$\dot{x}_5 = \frac{1}{T_{t2}}(x_6 - x_5) \tag{8}$$

$$\dot{x}_6 = \frac{1}{T_{sg2}}(u_2 - \frac{x_4}{R_2} - x_6) \tag{9}$$

$$\dot{x}_7 = 2\pi T_{12}(x_1 - x_4) \tag{10}$$

$$\dot{u}_1 = x_1e_1 + x_2e_2 + x_3e_3 + x_4e_4 + x_5e_5 + x_6e_6 + x_7e_7 + w_1e_8 \tag{11}$$

$$\dot{u}_2 = x_1f_1 + x_2f_2 + x_3f_3 + x_4f_4 + x_5f_5 + x_6f_6 + x_7f_7 + w_2e_8 \tag{12}$$

The detailed expressions of the coefficients $e$ and $f$ of Eq. (11-12) can be found in appendix C. The state matrix which is derived from Eq. (4-12) can be defined as,

$$A = \begin{bmatrix} \frac{-1}{T_{ps1}} & \frac{K_{ps1}}{T_{ps1}} & 0 & 0 & 0 & 0 & \frac{-K_{ps1}}{T_{ps1}} & 0 & 0 \\ 0 & \frac{-1}{T_{t1}} & \frac{1}{T_{t1}} & 0 & 0 & 0 & 0 & 0 & 0 \\ \frac{-1}{R_1 T_{sg1}} & 0 & \frac{-1}{T_{sg1}} & 0 & 0 & 0 & 0 & \frac{1}{T_{sg1}} & 0 \\ 0 & 0 & 0 & \frac{-1}{T_{ps2}} & \frac{K_{ps2}}{T_{ps2}} & 0 & \frac{K_{ps2}}{T_{ps2}}a_{12} & 0 & 0 \\ 0 & 0 & 0 & 0 & \frac{-1}{T_{t2}} & \frac{1}{T_{t2}} & 0 & 0 & 0 \\ 0 & 0 & 0 & \frac{-1}{R_2 T_{sg2}} & 0 & \frac{-1}{T_{sg2}} & 0 & 0 & \frac{1}{T_{sg2}} \\ 2\pi T_{12} & 0 & 0 & -2\pi T_{12} & 0 & 0 & 0 & 0 & 0 \\ e_1 & e_2 & e_3 & e_4 & e_5 & e_6 & e_7 & 0 & 0 \\ f_1 & f_2 & f_3 & f_4 & f_5 & f_6 & f_7 & 0 & 0 \end{bmatrix} \tag{13}$$

## 3 Optimization Techniques

In this section, brief overviews of GBO and EGBO are presented, explaining how each algorithm works in the search for the optimal parameters and how the algorithms are related to each other. More details of the subsequent techniques can be found in [29, 30].

### 3.1 Gradient-Based Optimizer

GBO is a novel meta-heuristic optimization algorithm that explores the search domain using Newton's method [30]. GBO operates with the assistance of two preeminent operators, namely Gradient Search Rule (GSR) and Local Escaping Operator (LEO).

The gradient search rule regulates the movement of vectors to better search in the feasible domain and gets better locations using the following equations:

$$GSR = randn \times \frac{2\Delta x \times x_i}{(y_{1_i} - y_{2_i} + \varepsilon)} \tag{14}$$

Where,

$$y_{1_i} = rand \times \left[ \frac{(u_{i+1} + x_i)}{2} + rand \times \Delta x \right] \tag{15}$$

$$y_{2_i} = rand \times \left[ \frac{(u_{i+1} + x_i)}{2} + rand \times \Delta x \right] \tag{16}$$

$$u_{i+1} = x_i - randn \times \frac{2\Delta x \times x_i}{(x_{worst} - x_{best} + \varepsilon)} + DM \tag{17}$$

Here, $u_{i+1}$ is the new GBO solution, *randn* is a random integer with a normal distribution, $x_{worst}$ and $x_{best}$ are the worst and best solutions acquired during the optimization process and $\varepsilon$ represents a small-scale number in the range [0, 0.1], which helps the equations to escape from getting zero values in the denominators.

A direction movement is denoted by the letter *DM*, which is written as,

$$DM = rand \times F_2 \times (x_{best} - x_i^{it})$$ (18)

Here, $F_2$ is an adaptive parameter and $it$ is the iteration number. $\Delta x$ is defined as the difference between two vectors, which is formulated as,

$$\Delta x = rand \times |\chi|$$ (19)

In which,

$$\chi = \frac{(x_{best} - x_i^{it}) + \eta}{2}$$ (20)

$$\eta = 2 \times rand \times \left( \left| \frac{x_{a_1}^{it} + x_{a_2}^{it} + x_{a_3}^{it} + x_{a_4}^{it}}{4} - x_i^{it} \right| \right)$$ (21)

Here, $a_1$, $a_2$, $a_3$, and $a_4$ denote four random numbers in the range of [1, $N$], in which, $N$ is the pre-defined population number.

In each iteration of the optimization process, GBO uses the GSR and $DM$ to generate a new solution. As a result, the new solution may be computed as follows:

$$X_{new_{1_i}}^{it} = x_i^{it} - GSR + DM$$ (22)

$$X_{new_{1_i}}^{it} = x_i^{it} - randn \times F_1 \times \frac{2\Delta x \times x_i}{(y_{1_i} - y_{2_i} + \varepsilon)} + DM$$ (23)

To improve its local search, GBO generates another new solution vector using the following relation,

$$X_{new_{2_i}}^{it} = x_{best} - randn \times F_1 \times \frac{2\Delta x \times x_m^{it}}{(y_{1_i}^{it} - y_{2_i}^{it} + \varepsilon)} + rand \times F_2 \times (x_{a_1}^{it} - x_{a_2}^{it})$$ (24)

$F_1$ and $F_2$ are defined as adaptive scale factors which are used to strike a balance between exploration and exploitation, as well as to discover auspicious regions in the search space.

Combining these two new solution vectors of Eq. (23-24), the solution vector for a certain iteration is obtained as,

$$x_i^{it+1} = r_1 \times (r_2 \times X_{new_{1_i}}^{it} + (1 - r_2) \times X_{new_{2_i}}^{it}) + (1 - r_1) \times X_{new_{3_i}}^{it}$$ (25)

Here, $r_1$ and $r_2$ denotes integers that are randomly chosen from the range [0, 1] and,

$$X_{new_{3_i}}^{it} = x_i^{it} - F_1 \times (X_{new_{1_i}}^{it} - X_{new_{2_i}}^{it})$$ (26)

In the GBO, LEO is employed to boost the capacity to escape local solutions. A suitable solution can be constructed with the help of the following pseudocode:

$$\begin{aligned}
&\text{if } rand < 0.5 \\
&\quad \text{if } rand < 0.5 \\
&\quad\quad X_{LEO}^{it} = X_i^{it+1} + XL \\
&\quad \text{else} \\
&\quad\quad X_{LEO}^{it} = x_{best} + XL \\
&\quad \text{end} \\
&\quad X_i^{it+1} = X_{LEO}^{it} \\
&\text{end}
\end{aligned} \tag{27}$$

Here,

$$\begin{aligned}
XL = {} & \mu_1 \times (\theta_1 \times x_{best} - \theta_2 \times x_r^{it}) + \mu_2 \times F_1 \times \left(\theta_3 \times \left(X_{new_{2m}^{it}} - X_{new_{1m}^{it}}\right)\right. \\
& \left. + \theta_2 \times \left(x_{a1}^{it} - x_{a2}^{it}\right)\right)/2
\end{aligned} \tag{28}$$

The $\mu_1$ and $\mu_2$ are two random integers that reside in the range of [-1, 1]. $\theta_1$, $\theta_2$, and $\theta_3$ are described as:

$$\theta_1 = \begin{Bmatrix} 2 \times rand & ; if\ rand_1 < 0.5 \\ 1 & ; otherwise \end{Bmatrix} \tag{29}$$

$$\theta_2 = \begin{Bmatrix} rand & ; if\ rand_1 < 0.5 \\ 1 & ; otherwise \end{Bmatrix} \tag{30}$$

$$\theta_3 = \begin{Bmatrix} rand & ; if\ rand_1 < 0.5 \\ 1 & ; otherwise \end{Bmatrix} \tag{31}$$

Here, $rand_1$ is a number in the range of [0, 1].

### 3.2 Enhanced Gradient-Based Optimizer

Properly setting F1 and F2 in the GBO is critical for improving precision and convergence speed. To increase the efficiency of the process in EGBO, small control parameter values are disproportionately assigned to superior solutions, whereas high values are disproportionately given to worse solutions [29]. A rank-based process is employed to accomplish this, in which all entities are sorted in ascending order depending on the values of their objective functions. Then, the adaptive parameters are constructed using the following pseudo-code,

$$\begin{aligned}
&\text{if } rand < 0.5 \\
&\quad F_1 = a_1/N + 0.1 \times randn \\
&\quad F_2 = a_2/N + 0.1 \times randn \\
&\text{else} \\
&\quad F_1 = i/N + 0.1 \times randn \\
&\quad F_2 = i/N + 0.1 \times randn \\
&\text{end}
\end{aligned} \tag{32}$$

The crossover operator is used to increase population variety by merging the GBO's solution vectors with the existing solution in the following way,

$$
\begin{aligned}
&\text{for } j = 1 \text{ to } d \\
&\quad \text{if } rand < pc_r \text{ or } j = j_{rand} \\
&\quad\quad \text{if } rand < 0.5 \\
&\quad\quad\quad X_{new_j} = X_{new_{1_j}} \\
&\quad\quad \text{else} \\
&\quad\quad\quad X_{new_j} = X_{new_{2_j}} \\
&\quad\quad \text{end} \\
&\quad \text{else} \\
&\quad\quad X_{new_j} = x_{i,j} \\
&\quad \text{end}
\end{aligned}
\tag{33}
$$

Here, $pc_r$ is defined as the crossover probability rate. $pc_r$ is defined based on the following equation,

$$
pc_r = (i/N) + 0.1 \times randn
\tag{34}
$$

A modified version of the LEO, named MLEO is created to improve the performance of the original LEO. The enhanced MLEO is presented in the pseudocode below,

$$
\begin{aligned}
&\text{if } rand < LC \\
&\quad \text{if } rand < 0.5 \times (1 - \frac{it}{MaxIt}) \\
&\quad\quad X_{new} = x_{best_3} + F_1 \times (x_{best_2} - x_r^{it}) + F_2 \times (x_{a_1}^{it} - x_{a_2}^{it}) \\
&\quad \text{else} \\
&\quad\quad X_{new} = x_{best} + F_1 \times (x_{best_2} - x_r^{it}) + F_2 \times (x_{a_1}^{it} - x_{a_2}^{it}) \\
&\quad \text{end} \\
&\text{end}
\end{aligned}
\tag{35}
$$

In this work, $LC$ denotes a chaotic logistic map, initiated with a value of 0.7. $LC$ is updated in each iteration following the equation below,

$$
LC = 4 \times LC \times (1 - LC)
\tag{36}
$$

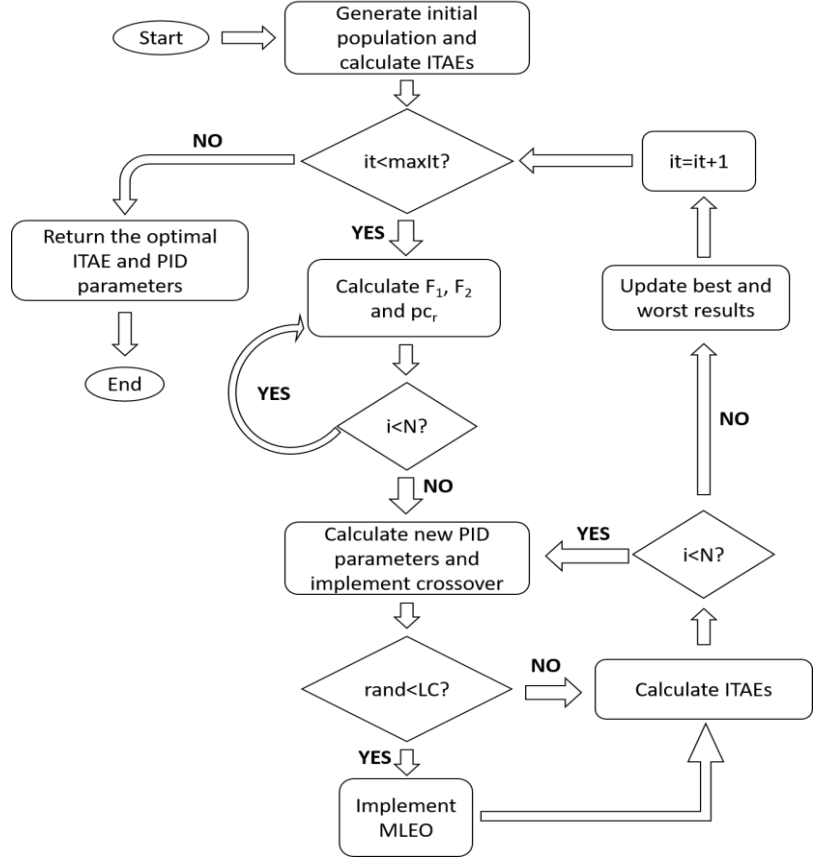

**Fig. 2.** Flowchart of EGBO based on the proposed controller

## 4 Controller Structure with Problem Formulation

In the LFC field, application of several kinds of controllers are observed such as PI controllers [16, 18], PID controllers [3–5, 15, 16, 20], PIDN controllers [13, 14], FOPID controllers [6, 11, 17, 19, 21, 28], FPID controllers [18, 23, 24], cascaded controllers [8–10], neural network-based controllers [3, 27], etc. For this study, a generic PID controller is chosen due to its industry-wide availability and affordability. The transfer function of the proposed controller, which is adopted from [33]:

$$TF_{PID} = K_p + \frac{K_i}{s} + K_d s \qquad (37)$$

$K_p$, $K_i$, and $K_d$ are the proportional, integral, and derivative gains of the controllers respectively. The inputs of the controllers are ACEs defined as,

$$ACE_1 = (K_{p1} + \frac{K_{i1}}{s} + K_{d1}s)u_1 = x_1b_1 + x_7 \tag{38}$$

$$ACE_2 = (K_{p2} + \frac{K_{i2}}{s} + K_{d2}s)u_2 = x_4b_2 - a_{12}x_7 \tag{39}$$

The operational objectives of the LFC are to maintain reasonably uniform frequency, to divide the load between generators, and to control the tie-line interchange schedules [2]. To keep the system frequency deviation within acceptable limits, the controller gains are optimized with the objective of minimization of ITAE, reducing ACEs of both areas, along with the deviation of the tie-line power flow. Hence, the optimization problem is formulated as follows:

Minimize,

$$ITAE = \int_{t=0}^{t_{final}} t(|x_1| + |x_4| + |x_7|)\,dt \tag{40}$$

subject to the following controller gain boundaries:

$$K_{p1}^{min} \le K_{p1} \le K_{p1}^{max} \tag{41}$$

$$K_{i1}^{min} \le K_{i1} \le K_{i1}^{max} \tag{42}$$

$$K_{d1}^{min} \le K_{d1} \le K_{d1}^{max} \tag{43}$$

$$K_{p2}^{min} \le K_{p2} \le K_{p2}^{max} \tag{44}$$

$$K_{i2}^{min} \le K_{i2} \le K_{i2}^{max} \tag{45}$$

$$K_{d2}^{min} \le K_{d2} \le K_{d2}^{max} \tag{46}$$

**Table 2.** The Search Space of the Controllers

| $K_{p1}^{min}, K_{p2}^{min}$ | $K_{p1}^{max}, K_{p2}^{max}$ | $K_{i1}^{min}, K_{i2}^{min}$ | $K_{i1}^{max}, K_{i2}^{max}$ | $K_{d1}^{min}, K_{d2}^{min}$ | $K_{d1}^{max}, K_{d2}^{max}$ |
|---|---|---|---|---|---|
| -16 | -6 | -45 | -15 | -8 | -3 |

## 5 Study Results and Discussions

For a comprehensive comparative analysis, along with EGBO, five other optimization algorithms, namely GBO, GWO, PSO, SCA, and ChOA are adopted in the PID controller gain optimization process. For the sake of fair competition, the number of population or search agents for each algorithm is chosen to be 100, with a maximum of 500 iterations. All of the algorithms are initialized with random PID gain parameters within the same boundary constraints as shown in Eq. (41-46) and table 2 and run for 30 times. Appendix D contains additional relevant parameters for the aforementioned algorithms, which are determined after rigorous trial and error runs.

### 5.1 Case Studies

For the breakdown of the performances of the controller optimized by the aforementioned algorithms, the following case studies are conducted:

- **Case-1:** 15% constant SLP in both areas
- **Case-2:** 25% constant SLP in area 1 and no SLP in area 2
- **Case-3:** 25% constant SLP in area 2 and no SLP in area 1
- **Case-4:** 20% constant SLP in area 1 and 10% constant SLP in area 2
- **Case-5:** 20% constant SLP in area 2 and 10% constant SLP in area 1

**Table 3.** Optimal Values of the Controller for Case-1

| Optimizer | Best Optimal ITAE | Exe. Time (sec) | Controller 1 Parameters | | | Controller 2 Parameters | | |
|---|---|---|---|---|---|---|---|---|
| | | | $K_{p1}$ | $K_{i1}$ | $K_{d1}$ | $K_{p2}$ | $K_{i2}$ | $K_{d2}$ |
| ChOA | 0.2344 | 94.3089 | -14.9522 | -41.6957 | -5.6072 | -14.7309 | -43.1649 | -5.618 |
| **EGBO** | **0.2292** | **74.8221** | -15.1838 | -43.5993 | -5.7641 | -15.1738 | -45 | -5.761 |
| GBO | 0.2293 | 76.1791 | -15.2024 | -43.5671 | -5.7899 | -15.178 | -44.9967 | -5.7857 |
| GWO | 0.2298 | 95.7718 | -15.1871 | -43.7646 | -5.7986 | -15.0911 | -45 | -5.7799 |
| PSO | 0.2296 | 81.6852 | -15.2116 | -43.837 | -5.8098 | -15.1177 | -45 | -5.7709 |
| SCA | 0.2359 | 97.6668 | -16 | -45 | -6.103 | -16 | -45 | -5.9294 |

**Case-1:** From table 3, it is evident that EGBO provides the best optimal ITAE while taking the least amount of time. In terms of both best ITAE and execution time, EGBO edges over second-placed GBO and third-placed PSO. In Fig. 3, although the algorithms show a similar trend in terms of FD, for the case of TPD, it is evident that EGBO is by far the superior algorithm, boasting very little overshoot and settling in around 6-second.

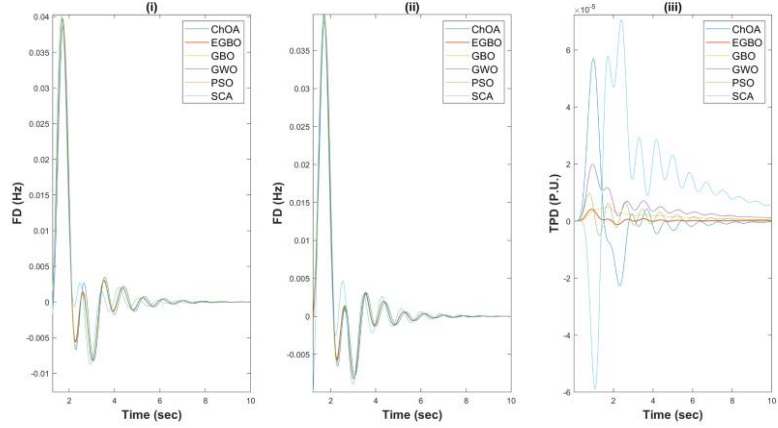

**Fig. 3.** (i) FD in area 1 (ii) FD in area 2 (iii) TPD for case-1

**Table 4.** Optimal Values of the Controller for Case-2

| Optimizer | Best Optimal ITAE | Exe. Time (sec) | Controller 1 Parameters | | | Controller 2 Parameters | | |
|---|---|---|---|---|---|---|---|---|
| | | | $K_{p1}$ | $K_{i1}$ | $K_{d1}$ | $K_{p2}$ | $K_{i2}$ | $K_{d2}$ |
| ChOA | 0.9384 | **95.4597** | -10.9494 | -45 | -4.7316 | -16 | -45 | -5.0483 |
| **EGBO** | **0.9378** | 98.3608 | -10.8375 | -45 | -4.7069 | -16 | -39.1146 | -4.8704 |
| GBO | **0.9378** | 100.6737 | -10.8375 | -45 | -4.7069 | -16 | -39.1148 | -4.8702 |
| GWO | 0.9379 | 96.9174 | -10.8392 | -45 | -4.7091 | -15.955 | -41.607 | -4.9248 |
| PSO | **0.9378** | 105.1054 | -10.8375 | -45 | -4.7069 | -16 | -39.1335 | -4.8708 |
| SCA | 0.939 | 100.0173 | -10.8832 | -45 | -4.7175 | -15.0735 | -21.4369 | -5.774 |

**Case-2:** For case-2, EGBO, GBO, and PSO deliver the same optimal ITAE of 0.9378, while GWO just falls short with an ITAE of 0.9379 as disclosed on table 4. Some of the controller parameters obtained from EGBO, GBO, and PSO are almost identical, which resonate with the simulation of FD in area 1 and TPD in Fig. 4. However, SLP applied on only area 1 seems to have an adverse effect on FD of area 2. The algorithms seem to find it difficult to minimize the oscillation and to minimize the settling time, albeit EGBO, which provides a smoother response with a faster settling time.

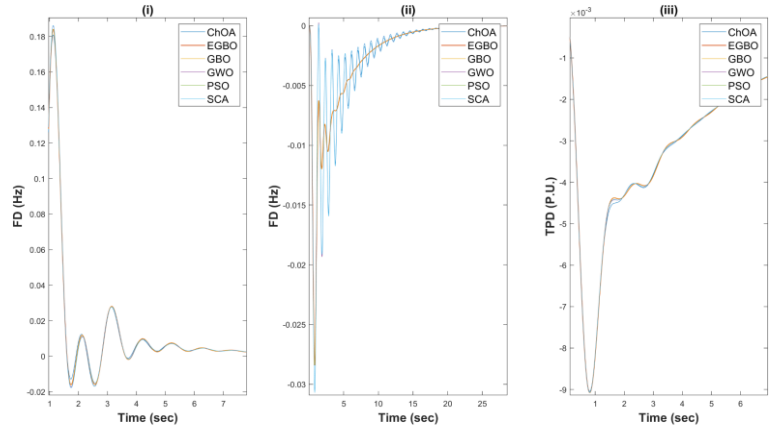

**Fig. 4.** (i) FD in area 1 (ii) FD in area 2 (iii) TPD for case-2

**Table 5.** Optimal Values of the Controller for Case-3

| Optimizer | Best Optimal ITAE | Exe. Time (sec) | Controller 1 Parameters | | | Controller 2 Parameters | | |
|---|---|---|---|---|---|---|---|---|
| | | | $K_{p1}$ | $K_{i1}$ | $K_{d1}$ | $K_{p2}$ | $K_{i2}$ | $K_{d2}$ |
| ChOA | 0.9507 | **97.3061** | -16 | -43.5345 | -5.0029 | -10.7151 | -45 | -4.6522 |
| **EGBO** | **0.95** | 99.9523 | -16 | -45 | -4.9366 | -10.6501 | -45 | -4.6468 |
| GBO | **0.95** | 105.6416 | -16 | -45 | -4.9366 | -10.6501 | -45 | -4.6468 |
| GWO | 0.9501 | 103.3018 | -16 | -45 | -4.9314 | -10.6555 | -45 | -4.6481 |
| PSO | **0.95** | 113.7967 | -16 | -45 | -4.9366 | -10.6501 | -45 | -4.6468 |
| SCA | 0.9528 | 101.2585 | -11.3121 | -45 | -5.0956 | -10.6818 | -45 | -4.6705 |

**Case-3:** For case-3, the aforesaid effect of only one area load perturbation on the FD and TPD simulations as explained for case-2 can be seen on Fig. 5, as well as similar trends for ITAE and PID gains on table 5. For further discussions on case-2 and case-3, the execution time can be accounted for. For both cases, ChOA exhibits the smallest execution times, but not the best optimal ITAE.

Among the algorithms that provide the best optimal ITAEs, EGBO converges to the solution, the quickest.

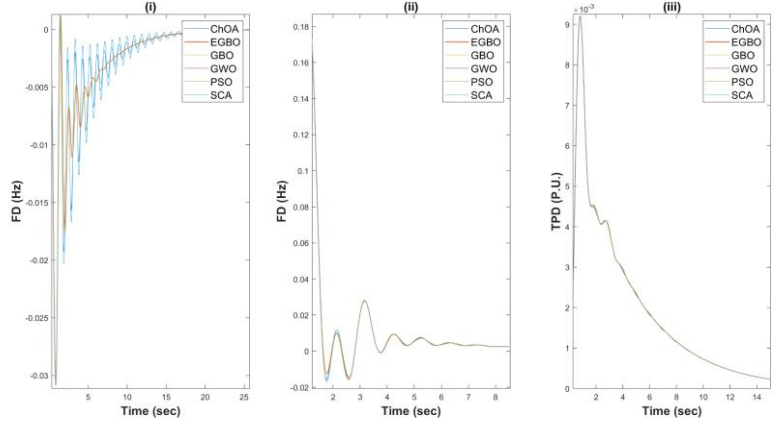

**Fig. 5.** (i) FD in area 1 (ii) FD in area 2 (iii) TPD for case-3

**Table 6.** Optimal Values of the Controller for Case-4

| Optimizer | Best Optimal ITAE | Exe. Time (sec) | Controller 1 Parameters | | | Controller 2 Parameters | | |
|---|---|---|---|---|---|---|---|---|
| | | | $K_{p1}$ | $K_{i1}$ | $K_{d1}$ | $K_{p2}$ | $K_{i2}$ | $K_{d2}$ |
| ChOA | 0.2805 | 93.7092 | -15.64 | -45 | -5.931 | -16 | -23.8575 | -5.6133 |
| **EGBO** | **0.2735** | **82.1557** | -15.292 | -45 | -5.856 | -15.3763 | -23.4416 | -5.4057 |
| GBO | 0.2736 | 82.6738 | -15.309 | -45 | -5.8478 | -15.4755 | -23.4111 | -5.382 |
| GWO | 0.2739 | 92.401 | -15.2249 | -45 | -5.8318 | -15.4385 | -23.404 | -5.3673 |
| PSO | 0.2737 | 85.5741 | -15.2937 | -45 | -5.863 | -15.425 | -23.4928 | -5.4229 |
| SCA | 0.2821 | 101.5411 | -16 | -45 | -5.9795 | -15.478 | -22.6949 | -5.3067 |

**Case-4:** From table 6, it is clear that EGBO possesses the best optimal ITAE value and execution time among the optimizers, surpassing its parent GBO by a slight margin. SCA shows better overshoot in the case of TPD as seen in Fig. 6, however, it requires a longer execution time.

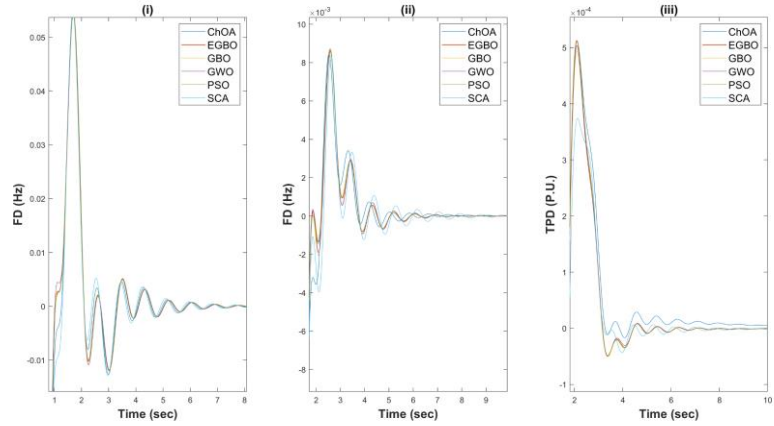

**Fig. 6.** (i) FD in area 1 (ii) FD in area 2 (iii) TPD for case-4

**Table 7.** Optimal Values of the Controller for Case-5

| Optimizer | Best Optimal ITAE | Exe. Time (sec) | Controller 1 Parameters | | | Controller 2 Parameters | | |
|---|---|---|---|---|---|---|---|---|
| | | | $K_{p1}$ | $K_{i1}$ | $K_{d1}$ | $K_{p2}$ | $K_{i2}$ | $K_{d2}$ |
| ChOA | 0.285 | 92.4471 | -15.0874 | -22.0303 | -5.4886 | -14.6593 | -45 | -5.7556 |
| **EGBO** | **0.2772** | **78.7314** | -15.1578 | -21.3754 | -5.3152 | -15.0066 | -45 | -5.3152 |
| GBO | 0.2773 | 79.3398 | -15.1262 | -21.3442 | -5.3058 | -15.0664 | -44.9952 | -5.7503 |
| GWO | 0.2775 | 93.1188 | -15.1036 | -21.3543 | -5.3137 | -15.0168 | -45 | -5.7608 |
| PSO | 0.2773 | 84.7991 | -15.1633 | -21.3759 | -5.3172 | -15.0122 | -45 | -5.7557 |
| SCA | 0.2892 | 97.3272 | -16 | -22.3174 | -5.519 | -15.3498 | -45 | -5.6841 |

**Case-5:** For case-5, the superiority of EGBO is crystal clear, which is evident from Fig. 7 as both FD and TPD simulations exhibit less jitter and faster settling time. Fig. 7 is the justification of the lowest optimal ITAE value provided by EGBO among the algorithms. EGBO is also the fastest to converge to such value as observed on table 7.

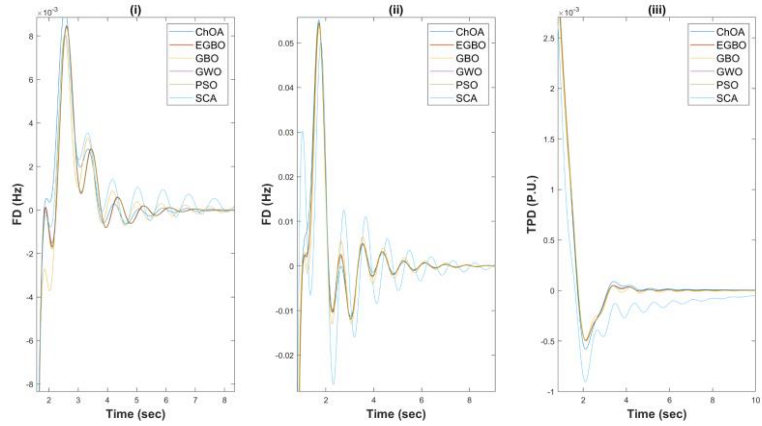

**Fig. 7.** (i) FD in area 1 (ii) FD in area 2 (iii) TPD for case-5

## 5.2 Convergence Comparison of the Optimizers

As explained earlier, it is vital to know exactly how fast the algorithm traverses the search space to get to the optimal fitness value. For an industry-wide application, the optimized controllers should be able to cope up the dynamic load, meaning, coping with the variation in SLP after a certain amount of time. For demonstrating the performance comparison, the first 50 iterations of case-5 are taken, along with the fitness value after 50 iterations, time taken for the algorithms to obtain such fitness value and the optimal ITAEs found in table 7. From Fig. 8, it can be seen that, although GBO performs well for the first few iterations, EGBO provides the lowest ITAE, which is apparent in table 8. For case-5, EGBO took around 80 seconds and 500 iterations to find the optimal ITAE. However, for the same case, after 50 iterations, EGBO possesses a fitness value of 0.2776, which is 0.14% closer to the optimal ITAE. To achieve this fitness value, EGBO takes only 8 seconds which is 10 times lower than as seen in table 7.

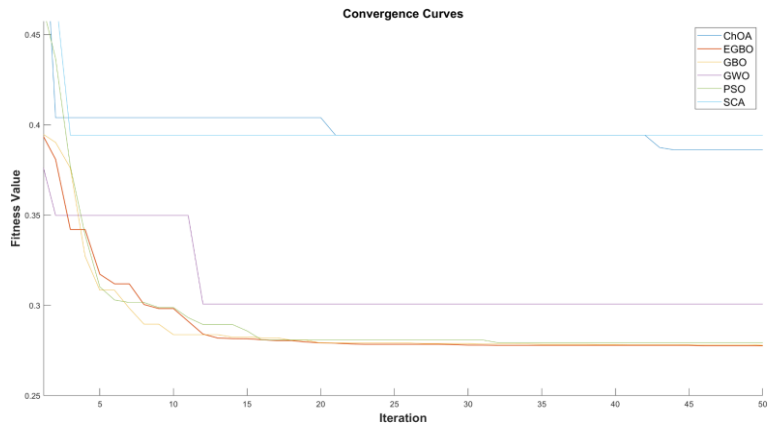

**Fig. 8.** Convergence curves of the algorithms under comparison

**Table 8.** Comparison of the Controllers after 50 iterations

| Optimizer | Fitness Value | Optimal ITAE | Exe. Time (sec) |
|---|---|---|---|
| ChOA | 0.3861 | 0.285 | 11.1407 |
| **EGBO** | **0.2776** | **0.2772** | **8.4618** |
| GBO | 0.2779 | 0.2773 | 8.9891 |
| GWO | 0.3007 | 0.2775 | 10.1287 |
| PSO | 0.2793 | 0.2773 | 8.8641 |
| SCA | 0.3943 | 0.2892 | 11.0746 |

From Fig. 8, it is discernible that ChOA and SCA are stuck into the local optima. For high-dimensional problems, ChOA has poor convergence and a tendency to become trapped in local minima [34]. SCA includes numerous search criteria to aid the search process, however, these characteristics might render the SCA vulnerable to local minima [35].

**Table 9.** Descriptive Statistical Study of the Algorithms for the Considered Five Cases

| Cases | Statistical Measures | ChOA | EGBO | GBO | GWO | PSO | SCA |
|---|---|---|---|---|---|---|---|
| **Case-1** | Mean | 0.2358 | **0.2296** | 0.23 | 0.2302 | 0.2303 | 0.2471 |
| | Best | 0.2344 | **0.2292** | 0.2293 | 0.2298 | 0.2296 | 0.2359 |
| | Worst | 0.2383 | **0.2301** | 0.2303 | 0.2304 | 0.2321 | 0.2673 |
| **Case-2** | Mean | 0.9395 | **0.9378** | 0.9379 | 0.9384 | 0.9379 | 0.94 |
| | Best | 0.9384 | **0.9378** | **0.9378** | 0.9379 | **0.9378** | 0.939 |
| | Worst | 0.9408 | **0.9378** | 0.9391 | 0.9393 | 0.9391 | 0.9416 |
| **Case-3** | Mean | 0.9544 | **0.95** | **0.95** | 0.9506 | **0.95** | 0.9575 |
| | Best | 0.9507 | **0.95** | **0.95** | 0.9501 | **0.95** | 0.9528 |
| | Worst | 0.9582 | **0.95** | 0.9501 | 0.9532 | 0.9501 | 0.9641 |
| **Case-4** | Mean | 0.2896 | **0.2737** | 0.2754 | 0.2741 | 0.2738 | 0.3021 |
| | Best | 0.2805 | **0.2735** | 0.2736 | 0.2739 | 0.2737 | 0.2821 |
| | Worst | 0.303 | **0.2738** | 0.2979 | 0.2743 | 0.2748 | 0.3222 |
| **Case-5** | Mean | 0.2961 | **0.2773** | 0.2792 | 0.2778 | 0.2784 | 0.3077 |
| | Best | 0.285 | **0.2772** | 0.2773 | 0.2775 | 0.2773 | 0.2892 |
| | Worst | 0.3088 | **0.2775** | 0.3043 | 0.278 | 0.3032 | 0.3215 |

### 5.3    Comparative Statistical Analysis

In this subsection, a comprehensive study is conducted to assess the statistical robustness of the optimized controllers. An important point to be noted is that all tests are performed at significance level of 0.05.

For each case, the algorithms are run for 30 times. The case study section of the paper is comprised of the best cases found while simulating the algorithms. For the comparative statistical analysis, all of the 30 runs are considered and the best, worst, and mean values of each of the algorithms for the five cases are calculated and presented in table 9. EGBO is the superior of the studied algorithms, performing efficiently all across the board. For case-2 and case-3, EGBO provides the same value for the considered three statistical measures, showcasing the robustness of the optimizer.

The two-way analysis of variance (ANOVA) test looks at the impact of several categorical independent variables on a single dependent variable, under following assumptions [36]:

‒  Independence of observation
‒  Normality
‒  Homoscedasticity

**Table 10.** Shapiro-Wilk Test of Normality

| Optimizers | Statistic | df | Sig. |
|------------|-----------|-----|------|
| ChOA | 0.6725 | 150 | 0 |
| EGBO | 0.6612 | 150 | 0 |
| GBO | 0.6617 | 150 | 0 |
| GWO | 0.6596 | 150 | 0 |
| PSO | 0.6597 | 150 | 0 |
| SCA | 0.6794 | 150 | 0 |

In our study, the occurrence of one observation has no bearing on the occurrence of the other observations, which satisfies the initial criteria. Table 11 shows that optimization methods have identical variance in terms of ITAE since the thresholds of significance are bigger than 0.05. In terms of normality, however, as evident on table 10, the null hypotheses are rejected ($p - values < 0.05$), therefore the two-way ANOVA test is skipped to increase the generality of the findings [37].

**Table 11.** Levene Test of Homogeneity of Variance

|  | Levene Statistic | df1 | df2 | Sig. |
|---|---|---|---|---|
| **Based on Mean** | 0.4989 | 5 | 894 | 0.7772 |
| **Based on Median** | 0.0086 | 5 | 894 | 1 |
| **Based on Median and with adjusted df** | 0.0086 | 5 | 892.16 | 1 |
| **Based on trimmed mean** | 0.4125 | 5 | 894 | 0.8403 |

**Friedman Test:** The Friedman test assumes no particular parametric form for the response distribution in each group, and the test does not require that the response distribution be normal [38]. Friedman's ranking test is examined in this study to test the premise that optimizing algorithms have a statistically significant influence on ITAE values.

**Table 12.** Ranks Based on Friedman Test

| Optimizers | Mean Rank |
|---|---|
| ChOA | 4.92 |
| EGBO | **1.94** |
| GBO | 2.01 |
| GWO | 3.73 |
| PSO | 2.42 |
| SCA | 5.98 |

The Friedman test analyzes the mean rankings of related groups and determines how they vary. As the *p-value* is less than the significant level as seen from table 13, the null hypothesis is rejected, inferring that not all group medians are equal. The test also renders a *chi-square* value of 605.0794, which further strengthens the notion that the optimizer type has a significant effect on ITAE, i.e., there is statistically compelling evidence of the optimizers being contrasting, cooperatively.

**Table 13.** Test Statistics of Friedman Analysis

| N | 150 |
|---|---|
| **Kendall's W** | 0.8068 |
| **Chi-Square** | 605.0794 |
| **df** | 5 |
| **Asymp. Sig.** | 0 |

Table 13 also encompasses the Kendall's coefficient of concordance result, which indicates a higher level of ranking agreement ($W = 0.8068$). As Friedman analysis rejects the null hypothesis, a pairwise comparison is utilized as a post-hoc analysis, which is essential to deem how each pair of optimizers are different from each other, statistically. The significance level is adjusted by the Bonferroni correction procedure, which is one of the most prominent approaches for controlling family-wise error rate, reducing type-I errors (false positives) [39].

**Table 14.** Pairwise Comparisons

| Sample1-Sample2 | Test Statistic | Std. Error | Std. Test Statistic | Sig. | Adj. Sig. |
|---|---|---|---|---|---|
| EGBO-GBO | -0.067 | 0.216 | -0.309 | 0.758 | **1** |
| GBO-PSO | -0.413 | 0.216 | -1.913 | 0.056 | **0.836** |
| EGBO-PSO | 0.48 | 0.216 | -2.222 | 0.026 | **0.394** |
| EGBO-GWO | -1.793 | 0.216 | -8.302 | 0.000 | 0.000 |
| EGBO-ChOA | 2.980 | 0.216 | 13.795 | 0.000 | 0.000 |
| ChOA-SCA | -1.060 | 0.216 | -4.907 | 0.000 | 0.000 |
| EGBO-SCA | -4.040 | 0.216 | -18.702 | 0.000 | 0.000 |
| GBO-ChOA | 2.913 | 0.216 | 13.486 | 0.000 | 0.000 |
| PSO-SCA | -3.560 | 0.216 | -16.480 | 0.000 | 0.000 |
| GWO-SCA | -2.247 | 0.216 | -10.400 | 0.000 | 0.000 |
| PSO-GWO | 1.313 | 0.216 | 6.080 | 0.000 | 0.000 |
| PSO-ChOA | 2.500 | 0.216 | 11.573 | 0.000 | 0.000 |
| GBO-GWO | -1.727 | 0.216 | -7.993 | 0.000 | 0.000 |
| GBO-SCA | -3.973 | 0.216 | -18.393 | 0.000 | 0.000 |
| GWO-ChOA | 1.187 | 0.216 | 5.493 | 0.000 | 0.000 |

Table 14 shows that, with the exception of (EGBO, GBO), (GBO, PSO), and (EGBO, PSO), all other pairings are statistically substantially different.

However, the problem with Bonferroni correction is that it is unduly conservative [39]. Bonferroni adjustments do not ensure a judicious interpretation of results as type-I errors cannot be reduced without inflating type-II errors (false negatives) [40].

**Wilcoxon Signed-Rank Test:** The Wilcoxon Signed-Rank analysis is applied to the aforementioned pairs, without any *p-value* adjustments. The rankings are provided in table 15, which imply that EGBO imparts lower ITAE values than GBO and PSO, for several independent simulations.

**Table 15.** Ranks Based on Wilcoxon-Signed Test

|  |  | N | Mean Rank | Sum of Ranks |
|---|---|---|---|---|
| **EGBO - GBO** | Negative Ranks | 80[a] | 83.46 | 6676.5 |
|  | Positive Ranks | 64[b] | 58.8 | 3763.5 |
|  | Ties | 6[c] |  |  |
|  | Total | 150 |  |  |
| **GBO - PSO** | Negative Ranks | 86[d] | 74.9 | 6441.5 |
|  | Positive Ranks | 52[e] | 60.57 | 3149.5 |
|  | Ties | 12[f] |  |  |
|  | Total | 150 |  |  |
| **EGBO - PSO** | Negative Ranks | 88[g] | 86.55 | 7616 |
|  | Positive Ranks | 58[h] | 53.71 | 3115 |
|  | Ties | 4[i] |  |  |
|  | Total | 150 |  |  |

a. EGBO <GBO
b. EGBO >GBO
c. EGBO = GBO
d. GBO <PSO
e. GBO >PSO
f. GBO = PSO
g. EGBO <PSO
h. EGBO >PSO
i. EGBO = PSO

**Table 16.** Test Statistics of Wilcoxon Signed-Rank Analysis

|  | EGBO - GBO | GBO - PSO | EGBO - PSO |
|---|---|---|---|
| **Z** | -2.905[b] | -3.498[b] | -4.397[b] |
| **Asymp. Sig. (two-tailed)** | 0.004 | 0 | 0 |

b. Based on positive ranks.

The *z-scores* are calculated by comparing the rank mean of each group to the total rank mean, which are -2.9047, -3.498, and -4.397 respectively, as presented in table 16. The *p-values* obtained from the analysis suggest that the null hypotheses can be rejected, which consequently deem EGBO, GBO, and PSO to be notably at odds, statistically.

# 6 Conclusion

This article describes the design and implementation of a new meta-heuristic algorithm, EGBO, to find an effective and optimal solution to the LFC problem in power systems. The PID parameters of a widely used two-area AGC system are optimized using EBGO, and later, evaluated. Other optimization algorithms including PSO, GWO, and SCA, as well as a few newer algorithms, such as GBO and ChOA, which are still uninvestigated in the LFC study, are employed to demonstrate the superiority of the proposed EGBO controller. A study encompassing five cases, employing SLP in both areas of the system is executed to assess the controller's robustness. Comparative statistical analysis is also conducted, along with a convergence comparison, to demonstrate how an EGBO optimized controller can perform in today's control system industries comprising of dynamic loads. This article opens the door to a new control mechanism to the LFC issue which is backed up by the performance of the algorithms for different SLPs and the statistical study of the algorithms for multiple independent runs, showcasing both stability and flexibility. EGBO can be the pioneer to the new wave of LFC problem studies, yielding the best possible optimal fitness values while taking the least amount of time.

# Appendix A

**Table 17.** Nomenclatures Used in the Studied System

| | |
|---|---|
| $R_1, R_2$ | Speed regulation constants of the governors of Area 1 and Area 2, respectively |
| $b_1, b_2$ | Frequency-bias factors of Area 1 and Area 2, respectively |
| $ACE_1, ACE_2$ | Area Control Errors of Area 1 and Area 2, respectively |
| $u_1, u_2$ | Outputs of the controllers of Area 1 and Area 2, respectively |
| $T_{sg1}, T_{sg2}$ | Time-constants for speed governor in seconds of Area 1 and Area 2, respectively |
| $x_3, x_6$ | Per-unit set position changes of governor valves of Area 1 and Area 2, respectively |
| $T_{t1}, T_{t2}$ | Time-constants for the turbine in seconds of Area 1 and Area 2, respectively |
| $x_2, x_5$ | Per-unit change in output powers of the turbines of Area 1 and Area 2, respectively |
| $T_{ps1}, T_{ps2}$ | Time-constants for the generator-load model in seconds of Area 1 and Area 2, respectively |
| $K_{ps1}, K_{ps2}$ | Gains of the generator-load model of Area 1 and Area 2, respectively |
| $w_1, w_2$ | A step-change in load demands of Area 1 and Area 2, respectively |
| $x_7$ | Tie-line power variation from its targeted value |
| $T_{12}$ | Coefficient of synchronization in seconds |
| $a_{12}$ | Constant |
| $x_1, x_4$ | Frequency deviations from their nominal value of Area 1 and Area 2, respectively |

# Appendix B

**Table 18.** Nominal Values Used in the Simulations

| $R_1, R_2$ | $b_1, b_2$ | $T_{sg1}, T_{sg2}$ | $T_{t1}, T_{t2}$ | $T_{ps1}, T_{ps2}$ | $K_{ps1}, K_{ps2}$ | $a_{12}$ |
|---|---|---|---|---|---|---|
| 3 | 0.425 | 0.4 | 0.5 | 20 | 100 | 1 |

## Appendix C

Various *e* and *f* coefficients of the studied system:

$$e_1 = 2\pi T_{12} K_{p1} + K_{i1} B_1 - \frac{K_{p1} B_1}{T_{ps1}} \tag{47}$$

$$e_2 = \frac{K_{p1} B_1 K_{ps1}}{T_{ps1}} - \frac{K_{d1} B_1 K_{ps1}}{T_{ps1} T_{t1}} - \frac{K_{d1} B_1 K_{ps1}}{T_{ps1}{}^2} + \frac{K_{d1} K_{ps1} 2\pi T_{12}}{T_{ps1}} \tag{48}$$

$$e_3 = \frac{K_{d1} B_1 K_{ps1}}{T_{ps1} T_{t1}} \tag{49}$$

$$e_4 = -K_{p1} 2\pi T_{12} + \frac{K_{d1} B_1 K_{ps1} 2\pi T_{12}}{T_{ps1}} + \frac{K_{d1} 2\pi T_{12}}{T_{ps2}} \tag{50}$$

$$e_5 = -\frac{K_{d1} K_{ps2} 2\pi T_{12}}{T_{ps2}} \tag{51}$$

$$e_6 = 0 \tag{52}$$

$$e_7 = K_{i1} - \frac{K_{p1} B_1 K_{ps1}}{T_{ps1}} + \frac{K_{d1} B_1 K_{ps1}}{T_{ps1}{}^2} - \frac{K_{d1} K_{ps1} 2\pi T_{12}}{T_{ps1}} - \frac{K_{d1} K_{ps2} 2\pi T_{12} a_{12}}{T_{ps2}} \tag{53}$$

$$e_8 = -\frac{K_{p1} B_1 K_{ps1}}{T_{ps1}} - \frac{K_{d1} 2\pi T_{12} K_{ps1}}{T_{ps1}} + \frac{K_{d1} B_1 K_{ps1}}{T_{ps1}{}^2} \tag{54}$$

$$e_9 = \frac{K_{d1} 2\pi T_{12} K_{ps2}}{T_{ps2}} \tag{55}$$

$$f_1 = -K_{p2} 2\pi T_{12} a_{12} + \frac{K_{d2} B_2 K_{ps2} 2\pi T_{12} a_{12}}{T_{ps2}} + \frac{K_{d2} a_{12} 2\pi T_{12}}{T_{ps1}} \tag{56}$$

$$f_2 = -\frac{K_{d2} K_{ps1} 2\pi T_{12} a_{12}}{T_{ps1}} \tag{57}$$

$$f_3 = 0 \tag{58}$$

$$f_4 = -\frac{K_{p2} B_2}{T_{ps2}} + K_{p2} 2\pi T_{12} a_{12} + K_{i2} B_2 - \frac{K_{d2} B_2 K_{ps2} 2\pi T_{12} a_{12}}{T_{ps2}} + \frac{K_{d2} B_2}{T_{ps2}{}^2} - \frac{K_{d2} 2\pi T_{12} a_{12}}{T_{ps2}} \tag{59}$$

$$f_5 = \frac{K_{p2} B_2 K_{ps2}}{T_{ps2}} - \frac{K_{d2} B_2 K_{ps2}}{T_{ps2} T_{t2}} - \frac{K_{d2} B_2 K_{ps2}}{T_{ps2}{}^2} + \frac{K_{d2} K_{ps2} 2\pi T_{12} a_{12}}{T_{ps2}} \tag{60}$$

$$f_6 = \frac{K_{d2} B_2 K_{ps2}}{T_{ps2} T_{t2}} \tag{61}$$

$$f_7 = \frac{K_{p2} B_2 K_{ps2} a_{12}}{T_{ps2}} - \frac{K_{d2} B_2 K_{ps2} a_{12}}{T_{ps2}{}^2} + \frac{K_{d2} K_{ps1} 2\pi T_{12} a_{12}}{T_{ps1}} - K_{i2} a_{12} + \frac{K_{d2} K_{ps2} 2\pi T_{12} a_{12}{}^2}{T_{ps2}} \tag{62}$$

$$f_8 = \frac{K_{d2} 2\pi T_{12} K_{ps1} a_{12}}{T_{ps1}} \tag{63}$$

$$f_9 = -\frac{K_{p2} B_2 K_{ps2}}{T_{ps2}} - \frac{K_{d2} 2\pi T_{12} K_{ps2} a_{12}}{T_{ps2}} + \frac{K_{d2} B_2 K_{ps2}}{T_{ps2}{}^2} \tag{64}$$

# Appendix D

**Table 19.** Parametric Values Used in the Optimizers

| Optimizer | Values of the control parameters |
|---|---|
| ChOA | SearchAgents_no = 100, Max_iteration = 500, dim = 6 |
| EGBO | nP = 100, MaxIt = 500, dim = 6, LC = 0.7 |
| GBO | nP = 100, MaxIt = 500, dim = 6, pr = 0.5 |
| GWO | SearchAgents_no = 100, Max_iteration = 500, dim = 6 |
| PSO | noP = 100, maxIter = 500, nVar = 6, wMax = 0.9, wMin = 0.2, c1 = 2, c2 = 2 |
| SCA | SearchAgents_no = 100, Max_iteration = 500, dim = 6, a = 2 |

## References

[1] Theodore Wildi. *Electrical machines, drives, and power systems*. Pearson Educaci´on, 2006.

[2] Hadi Saadat. "Power System Analysis,(2nd)". In: *McGraw-Hill Higher Education* (2009).

[3] Amin Safari, Farshad Babaei, and Meisam Farrokhifar. "A load frequency control using a PSO-based ANN for micro-grids in the presence of electric vehicles". In: *International Journal of Ambient Energy* 42.6 (2021), pp. 688–700.

[4] Hesam Parvaneh et al. "Load frequency control of a multi-area power system by optimum designing of frequency-based PID controller using seeker optimization algorithm". In: *2016 6th Conference on Thermal Power Plants (CTPP)*. IEEE. 2016, pp. 52–57.

[5] Anil Annamraju and Srikanth Nandiraju. "Coordinated control of conventional power sources and PHEVs using jaya algorithm optimized PID controller for frequency control of a renewable penetrated power system". In: *Protection and Control of Modern Power Systems* 4.1 (2019), pp. 1–13.

[6] Kumar Saurabh, Neelesh Kumar Gupta, and Arun Kumar Singh. "Fractional Order Controller Design for Load Frequency Control of Single Area and Two Area System". In: *2020 7th International Conference on Signal Processing and Integrated Networks (SPIN)*. IEEE. 2020, pp. 531–536.

[7] Naladi Ram Babu and Lalit Chandra Saikia. "Load Frequency Control of a Multi-area System Incorporating Dish-Stirling Solar Thermal System and Coyote Optimized PI minus DF Controller". In: *2020 IEEE International Conference on Power Electronics, Smart Grid and Renewable Energy (PESGRE2020)*. IEEE. 2020, pp. 1–6.

[8] Chittaranjan Pradhan and Terje Gjengedal. "Adaptive Jaya Algorithm for Optimized PI-PD Cascade Controller of Load Frequency Control in Interconnected Two-Area Power System". In: *2020 International Conference on Smart Systems and Technologies (SST)*. IEEE. 2020, pp. 181–186.

[9] Nimai Charan Patel et al. "Two-staged (PDF+ 1PI) controller design for load frequency control". In: *2020 International Conference on Computational Intelligence for Smart Power System and Sustainable Energy (CISPSSE)*. IEEE. 2020, pp. 1–6.

[10] Debasis Tripathy et al. "Grasshopper Optimization Algorithm based Fuzzy PD-PI Cascade Controller for LFC of Interconnected Power System Coordinate with Renewable sources". In: *2020 IEEE Calcutta Conference (CALCON)*. IEEE. 2020, pp. 111–116.

[11] Hao Chen et al. "Fractional-order PID Load Frequency Control for Power Systems Incorporating Thermostatically Controlled Loads". In: *2021 IEEE 4th International Electrical and Energy Conference (CIEEC)*. IEEE. 2021, pp. 1–6.

[12] Yuemin Zheng et al. "A novel chaotic fractional-order beetle swarm optimization algorithm and its application for load-frequency active disturbance rejection

control". In: *IEEE Transactions on Circuits and Systems II: Express Briefs* (2021).

[13] Saswati Mishra, Shubhrata Gupta, and Anamika Yadav. "Design and application of controller based on sine-cosine algorithm for load frequency control of power system". In: *International Conference on Intelligent Systems Design and Applications*. Springer. 2018, pp. 301–311.

[14] Sourabh Dewangan, Tapan Prakash, and Vinay Pratap Singh. "Design and performance analysis of elephant herding optimization based controller for load frequency control in thermal interconnected power system". In: *Optimal Control Applications and Methods* 42.1 (2021), pp. 144–159.

[15] B Vedik et al. "Renewable Energy-Based Load Frequency Stabilization of Interconnected Power Systems Using Quasi-Oppositional Dragonfly Algorithm". In: *Journal of Control, Automation and Electrical Systems* 32.1 (2021), pp. 227–243.

[16] Dipayan Guha, Provas Kumar Roy, and Subrata Banerjee. "Load frequency control of interconnected power system using grey wolf optimization". In: *Swarm and Evolutionary Computation* 27 (2016), pp. 97–115.

[17] Swati Sondhi and Yogesh V Hote. "Fractional order PID controller for perturbed load frequency control using Kharitonov's theorem". In: *International Journal of Electrical Power & Energy Systems* 78 (2016), pp. 884– 896.

[18] Mohammad Hassan Khooban et al. "A new load frequency control strategy for micro-grids with considering electrical vehicles". In: *Electric Power Systems Research* 143 (2017), pp. 585–598.

[19] Deepak Kumar Lal and Ajit Kumar Barisal. "Combined load frequency and terminal voltage control of power systems using moth flame optimization algorithm". In: *Journal of Electrical Systems and Information Technology* 6.1 (2019), pp. 1–24.

[20] Ankur Rai and Dushmanta Kumar Das. "Class topper optimization (cto) algorithm based load frequency control of multi-area interconnected power system with non-linearity". In: *2020 IEEE International Conference on Computing, Power and Communication Technologies (GUCON)*. IEEE. 2020, pp. 45–50.

[21] Andrew Xavier Raj Irudayaraj et al. "A Matignon's theorem based stability analysis of hybrid power system for automatic load frequency control using atom search optimized FOPID controller". In: *IEEE Access* 8 (2020), pp. 168751–168772.

[22] Erdinc Sahin. "Design of an optimized fractional high order differential feedback controller for load frequency control of a multi-area multi-source power system with nonlinearity". In: *IEEE Access* 8 (2020), pp. 12327– 12342.

[23] Gonggui Chen et al. "An improved ACO algorithm optimized fuzzy PID controller for load frequency control in multi area interconnected power systems". In: *IEEE Access* 8 (2019), pp. 6429–6447.

[24] Benazeer Begum et al. "Optimal design and implementation of fuzzy logic based controllers for LFC study in power system incorporated with wind firms". In:

*2020 International Conference on Computational Intelligence for Smart Power System and Sustainable Energy (CISPSSE)*. IEEE. 2020, pp. 1–6.

[25] Ch Naga Sai Kalyan et al. "Ascertainment of Appropriate GRC Structure for Two Area Thermal System under Seagull Optimization based 2DOF-PID Controller". In: *2021 7th International Conference on Signal Processing and Intelligent Systems (ICSPIS)*. IEEE. 2021, pp. 01–05.

[26] Emad M Ahmed et al. "Frequency regulation of electric vehicle-penetrated power system using MPA-tuned new combined fractional order controllers". In: *IEEE Access* 9 (2021), pp. 107548–107565.

[27] Milton Kumar Das et al. "PI-RLNN Controller for LFC of Hybrid Deregulated Power System Based on SPOA". In: *2021 IEEE 18th India Council International Conference (INDICON)*. IEEE. 2021, pp. 1–5.

[28] Chandresh Singh and Prabin Kumar Padhy. "Fractional Order Controller Design for interconnected Power System using BAT optimization Algorithm." In: *2022 Second International Conference on Artificial Intelligence and Smart Energy (ICAIS)*. IEEE. 2022, pp. 1634–1639.

[29] Iman Ahmadianfar et al. "Gradient-based optimization with ranking mechanisms for parameter identification of photovoltaic systems". In: *Energy Reports* 7 (2021), pp. 3979–3997.

[30] Iman Ahmadianfar, Omid Bozorg-Haddad, and Xuefeng Chu. "Gradientbased optimizer: A new metaheuristic optimization algorithm". In: *Information Sciences* 540 (2020), pp. 131–159.

[31] M Khishe and Mohammad Reza Mosavi. "Chimp optimization algorithm". In: *Expert systems with applications* 149 (2020), p. 113338.

[32] Dwarkadas Pralhaddas Kothari and IJ Nagrath. *Modern power system analysis*. Tata McGraw-Hill Education, 2003.

[33] *Book: Introduction to Control Systems (Iqbal)*. [Online; accessed 2022-0101]. 2020.

[34] Ch Kumari et al. "A boosted chimp optimizer for numerical and engineering design optimization challenges". In: *Engineering with Computers* (2022), pp. 1–52.

[35] Muhammad Zubair Rehman et al. "A new Multi Sine-Cosine algorithm for unconstrained optimization problems". In: *Plos one* 16.8 (2021), e0255269.

[36] George W Snedecor and William G Cochran. "Statistical Methods Iowa State University Press, Ames". In: *Statistical methods, 7th ed.. The Iowa State University Press, Ames* (1980).

[37] Dulce G Pereira, Anabela Afonso, and F´atima Melo Medeiros. "Overview of Friedman's test and post-hoc analysis". In: *Communications in StatisticsSimulation and Computation* 44.10 (2015), pp. 2636–2653.

[38] Roy St. Laurent and Philip Turk. "The effects of misconceptions on the properties of Friedman's test". In: *Communications in Statistics-Simulation and Computation* 42.7 (2013), pp. 1596–1615.

[39]  Sangseok Lee and Dong Kyu Lee. "What is the proper way to apply the multiple comparison test?" In: *Korean journal of anesthesiology* 71.5 (2018), p. 353.

[40]  Thomas V Perneger. "What's wrong with Bonferroni adjustments". In: *Bmj* 316.7139 (1998), pp. 1236–1238.